
\raggedbottom


\def\nex{\par\noindent\hang}

\null
\centerline{\bf ARE THE NEARBY AND PHOTOMETRIC}
\vskip 2mm
\centerline{\bf STELLAR LUMINOSITY FUNCTIONS DIFFERENT?
}
\vskip 10mm
\par
\centerline{\bf Pavel Kroupa}
\vskip 10mm
\centerline{Astronomisches Rechen-Institut}
\vskip 2mm
\centerline{M{\"o}nchhofstra{\ss}e~12-14, D-69120~Heidelberg, Germany}
\vskip 15mm
\centerline{e-mail: S48 @ IX.URZ.UNI-HEIDELBERG.DE}


\vskip 140mm

\centerline{ApJ, in press} (due in Nov.~1 issue of Part~1)

\vfill\eject

\centerline{\bf Abstract}
\vskip 5mm
\noindent
The stellar luminosity function derived from the sample of stars within
5.2--20~pc is the nearby luminosity function. The luminosity function obtained
from deep low spatial resolution surveys to distances of typically 100--200~pc
is the photometric
luminosity function. We obtain and tabulate a best estimate of the parent
distribution from which Malmquist corrected photometric luminosity functions
are sampled. This is presently the most {\it accurate} and {\it precise}
available estimate of the true distribution of stars with absolute
magnitude for $M_{\rm V}\le16.5$, as derived from low-resolution surveys. Using
rigorous statistical
analysis we find that the hypothesis that the nearby and photometric
luminosity functions estimate the same parent luminosity function can be
discarded with 99~per~cent confidence or better for $M_{\rm V}>13,M_{\rm
bol}>10$.

\vskip 5mm
\noindent {\it Subject headings:} stars: low-mass -- stars: luminosity
function

\vfill\eject

\vskip 0.1in
\vskip 24pt
\noindent{\bf 1 INTRODUCTION}
\vskip 12pt
\noindent
The difference between the faint end of the stellar luminosity function
estimated from the stellar sample in the immediate neighbourhood of the Sun and
the
luminosity function obtained from photographic surveys which reach to distances
of 100--200~pc has been discussed by Dahn, Liebert \& Harrington (1986).
They point out that the nearby luminosity function contains more than 2~sigma
more stars in the $M_{\rm V}=16-17$ double bin than the photometric luminosity
function estimated by Reid \& Gilmore (1982).
Since then there has been a substantial debate
on this question (see e.g. Stobie, Ishida \& Peacock 1989; Henry \&
McCarthy 1990; Kroupa, Tout \& Gilmore 1991; Reid 1991; Kroupa, Tout \&
Gilmore 1993) fuelled by the
discrepant conclusions made by researchers hampered by small number statistics.

Tinney (1994), however, concludes that ``the debate in
recent years over the `difference' between the photometrically- and
trigonometrically-selected LFs has been somewhat of a mare's nest''. He bases
this conclusion on the extensive photographic survey reported in Tinney, Reid
\& Mould (1993)
which estimates the photometric luminosity function to an unprecendented {\it
precision} because
of the very large sample size, being based on 3538~stars in the magnitude
range $8.5\le M_{\rm bol}\le 15.0$. Tinney (1993, 1994) and Reid
(1994) find that on the bolometric magnitude scale
there is no difference between the photometric luminosity function and the
nearby luminosity function at faint magnitudes.

It is because there appear in the recent literature contradictory opinions and
views as to the significance of, reality of and reason for the difference
between the nearby and photometric luminosity function
that we here cast a critical eye on the
observational data obtained by a number of researchers in order to (i)
establish the significance of the difference between the nearby and
photometric luminosity function, and (ii) to verify the
results obtained by Tinney (1993) and to study the {\it accuracy} of his
estimate of the photometric stellar luminosity function. We
are able
to do this by a statistical analysis of various observational estimates of the
photometric luminosity function which were not accessible to Dahn et al.
(1986).

We proceed as follows: We first show that the four independent observational
estimates of the Malmquist uncorrected photometric luminosity function by Reid
\& Gilmore (1982), Gilmore, Reid \& Hewett (1985), Stobie et al.
(1989) and Kirkpatrick et al. (1994) can be used to estimate the parent
Malmquist uncorrected photometric luminosity function. We then apply Malmquist
corrections and derive a best estimate of the true distribution of stars with
absolute V-band magnitude as obtained from photographic, or generally deep-sky,
surveys.
In Section~2 we introduce and tabulate the primary observational data we are
concerned with, and in Section~3 we estimate the parent distribution.
In Section~4 we test the hypothesis that the nearby
luminosity function estimates the same parent distribution as the best
estimate of the true
photometric luminosity function, and in Section~5 we show that the results
carry over into bolometric magnitudes and we compare our best estimate for
the true photometric luminosity function with the estimate used by Tinney
(1993). Section~6 contains our conclusions.

The possible reasons for the difference are discussed in Kroupa (1995a).

\vskip 0.1in
\vskip 24pt
\noindent{\bf 2 THE OBSERVATIONAL ESTIMATES}
\vskip 12pt
\noindent
To estimate the stellar luminosity function from
photographic surveys requires identifying all
low-mass (i.e. red) stars on the survey photographic plates and estimating
stellar number
densities using photometric parallax. This is a formidable task because the
significant contamination by galaxies, giant stars in the Galaxy and white
dwarfs needs to be eliminated. Modern automatic plate scanning machines
(e.g. COSMOS and APM, UK) enable this task to be performed.
Although only a narrow cone is available
the sampling volume in which star counts are complete is large owing to the
large sampling distance (100--200~pc).

Photographic star counts need to be
corrected for Malmquist bias (see Stobie et al. 1989 for a
thorough discussion) which arises because 1) mean absolute magnitudes of stars
of a particular colour in a magnitude limited sample are brighter than the mean
absolute magnitude in a volume limited sample, and 2) the stellar number
density in a magnitude limited sample is larger than in a volume limited
sample. Both effects arise because i) photometry has errors, and stars of a
given colour or mass have a
range of luminosities since they have different metallicities, ages, and may be
unresolved binary systems (this scatter in luminosities is referred to as
cosmic scatter), and ii) the number of stars increases nonlinearly with
increasing distance. Thus, intrinsically bright stars are
overrepresented in a magnitude limited sample.
We return to Malmquist corrections in Section~3.3. We refer to the
Malmquist uncorrected photometric luminosity function as the `raw'
or `observed' photometric luminosity function, $\Psi_{\rm phot}^*$. The
photometric
luminosity function corrected for Malmquist bias is $\Psi_{\rm phot}$.

Reid \& Gilmore (1982) published the first estimate of the faint
photometric stellar luminosity function, $\Psi_{\rm RG}^*(M_{\rm V})$,
obtained from a survey of photographic plates
in the direction of the South Galactic
Pole ($l=0,b=-90^{\rm o}$). They use the V- and I-band for photometric
parallax estimation and count 85 stars in the magnitude interval $8.5\le
M_{\rm V}\le 16.5$. We focuss our attention
on two further photographic surveys using the same photometric bands (Reid
(1991) comments on the much larger Malmquist corrections necessary when using
the steeper $M_{\rm I},R-I$ relation). These are
the surveys by Gilmore et al. (1985) in the direction $l=37^{\rm
o},b=-51^{\rm o}$, $\Psi_{\rm GRH}^*(M_{\rm V})
$, based on 64 stars in the magnitude interval $8.5\le M_{\rm V}\le 16.5$ and
by Stobie et al. (1989) towards
the North Galactic Pole ($l=0,b=+90^{\rm o}$), $\Psi_{\rm SIP}^*(M_{\rm V})$,
based on 178~stars in the magnitude interval $7.5\le M_{\rm V}\le
16.5$. We also use the recent V-, R- and I-band CCD Transit Instrument survey
of faint stars by Kirkpatrick et al. (1994). Their estimate of the photometric
luminosity function, $\Psi_{\rm Kirk}^*(M_{\rm V})$, is based on photometric
parallax estimation in the V- and I-bands and has~121 stars in
the magnitude interval $10.5\le M_{\rm V}\le 15.5$. The estimates of the
photometric luminosity function by Hawkins \& Bessell (1988) and Leggett \&
Hawkins (1988) agree with the other surveys (see Stobie et al. 1989 and
Krikpatrick et al. 1994, respectively) but we do not use these because their
estimates are obtained using photometric parallax in the R- and I-bands (see
Section~3.3).

Stobie et al. (1989) show that the
Malmquist corrections applied to the luminosity function prior to their study
were incomplete [because only point 1) above was corrected for, i.e. Reid \&
Gilmore (1982) and Gilmore et al. (1985) do not correct for the apparent
increase in stellar number density], and that the peak in the photometric
luminosity
function at $M_{\rm V}\approx12$ is much more enhanced after correction
for the full Malmquist bias. In particular, raw photometric luminosity
functions overestimate stellar number densities at $M_{\rm V}>12$.
The contribution of unresolved binaries and metallicity to cosmic
scatter varies with magnitude which would require an even more elaborate
treatment of Malmquist corrections than performed by Stobie et al.
who assumed a constant Gaussian cosmic scatter. Kirkpatrick et al. (1994)
account for a varying uncertainty in absolute magnitudes as a function of
absolute mangitude, but only consider measurement errors in their photometry,
$\sigma_{\rm phot}\approx0.1$~mag for $10.25<M_{\rm V}<14.00$. This severely
underestimates Malmquist bias
because it is proportional to $\sigma_{\rm tot}^2$ (Stobie et al. 1989).
The total scatter in absolute magnitudes owing to the
dispersion in metallicities, unresolved binary systems, ages and photometry
errors is $\sigma_{\rm tot}\approx0.5$~mag (Kroupa et al. 1993).
Summarising, we find that of all four estimates of the photometric luminosity
function only Stobie et al. (1989) have correctly applied Malmquist
corrections. To obtain an improved estimate of the parent distribution from
which Malmquist corrected photometric luminosity functions are sampled we
use the readily available raw luminosity functions to first improve our
estimate of the Malmquist
uncorrected parent distribution which we correct for Malmquist bias in
Section~3.3.

The sample of stars in the neighbourhood of
the sun allows estimation of the stellar luminosity function using
trigonometric parallax measurements. For $M_{\rm V}<11.5$ the complete sampling
distance
is 10--20~pc (Wielen, Jahreiss \& Kr{\"u}ger 1983), but for fainter
magnitudes one must resort to the stellar
sample within a distance of about 5~pc to obtain probably complete number
density estimates.
For $M_{\rm V}\ge11.5$ we base the nearby luminosity function, $\Psi_{\rm
near}$, on the stellar sample northward of declination $-20^{\rm o}$ and within
the 5.2~pc sampling distance (Dahn et al.
1986) extended in Kroupa et al. (1993) to include the discovery by
Henry \& McCarthy
(1990) that one of the `stars' in the 5.2~pc sample is a binary system.
The sampling volumes for $\Psi_{\rm near}$ are
$V=33510.3\,$pc$^3$ ($M_{\rm V}<7.5$), $V=25146.9$~pc$^3$
($7.5\le M_{\rm V}<9.5$),
$V=3143.4$~pc$^3$ ($9.5\le M_{\rm V}<11.5$) and $V=395.2$~pc$^3$
($11.5\le M_{\rm V}$).

Owing  to the small sampling distance the
nearby luminosity function is less well constrained for $M_{\rm V}>12$ than the
photometric luminosity function.
We would need an increase of the nearby sample
by at least an order of magnitude to {\it significantly} improve our estimate
of the nearby luminosity function. The parallax limited nearby star count data
suffer under
a bias similar to the Malmquist bias because of the finite error in parallax
measurements (Lutz \& Kelker 1973) and because the metallicity and age
dispersion smears out the shape of the nearby luminosity function
(compare figs.~1 and~20
in Kroupa et al. 1993). Corrections similar to the Malmquist
corrections have not been performed on the nearby sample to date. The results
of this paper are, however, not affected by neglecting to correct for
these effects because
the corrections are smaller than the statistical uncertainties per
magnitude bin which are very large for $M_{\rm V}>12$.

In Table~1 we tabulate the four estimates of the photometric luminosity
function not corrected for Malmquist bias by the respective authors.
Column~1 lists the absolute visual magnitude of the bin centre.
Columns~2 and~3 tabulate the raw
estimate of the photometric luminosity function by Reid \& Gilmore (1982),
$\Psi_{\rm RG}^*$ and the
number of stars per magnitude bin, $n_{\rm b}$, respectively. We have corrected
log$_{10}(\Psi_{\rm RG}^*)$ for
the disk density gradient of $+0.11$ as suggested by Reid \& Gilmore (1982),
but have ignored their $+0.08$~mag Malmquist correction. Similarly, Columns~4
and~5 contain $\Psi_{\rm GRH}^*$, Columns~6 and~7 tabulate
$\Psi_{\rm SIP}^*$, and Columns~8 and~9 list $\Psi_{\rm Kirk}^*$.
Columns~10 and~11 contain our best estimate of the
raw photometric luminosity function obtained in Section~3.2,
$\overline{\Psi}_{\rm phot}^*$.
In Table~2 we tabulate our best estimate of the Malmquist corrected photometric
luminosity function, $\overline{\Psi}_{\rm phot}$ (Columns~2 and~3, derived
in Section~3.3)  and
Columns~4 and~5 list the nearby luminosity function, $\Psi_{\rm near}$.
We note that in all cases the statistical uncertainty in $\Psi$ is
$\sigma=\Psi/n_{\rm b}^{1\over2}$.
The luminosity functions are accurate to three significant figures only, but we
list four to reduce rounding errors in our statistical tests. The four
observational estimates of the raw photometric and of the nearby luminosity
function are compared in Fig.~1.

It is quite possible that further faint stars
will be found within the 5.2~pc sampling distance. However,
we do not expect a significant revision of the stellar number density at
$M_{\rm V}<16$.
Recent trigonometric parallax measurements suggest that the distances to GL~445
and GJ~1116AB are 5.22 and 5.23~pc, respectively (T. Henry, private
communication). These stars are in the 5.2~pc sample of Kroupa et al. (1993),
from which the nearby luminosity function in Table~2 is estimated. However,
changing the nearby luminosity function to account for the nearby parallax
measurements is not warranted at this stage because: (i) The trigonometric
parallax measurement uncertainties imply an uncertainty in distance of about
0.1~pc at a distance of 5.2~pc so that both stars cannot be proven to lie
outside the 5.2~pc distance limit. (ii) A new sampling volume can be defined:
declination northward of $-20^{\rm o}$ as used by Kroupa et al. (1993), but
with a distance limit of 5.23~pc. This change in sampling volume would retain
the stars in the sample of Kroupa et al. (1993) but would change the
normalisation of the nearby luminosity function in Table~2 by 1.7~per cent
which is insignificant. A new star, GL~866C ($M_{\rm V}=15.8$) has been added
to the 5.2~pc sample (T. Henry, private communication). This would increase the
number density near $M_{\rm V}=16$ increasing the difference between the nearby
and photometric luminosity functions. By not counting this star we thus
underestimate the difference between the nearby and photometric luminosity
functions.

\vskip 24pt
\noindent{\bf 3 THE BEST ESTIMATE PHOTOMETRIC LUMINOSITY FUNCTION}
\vskip 12pt
\noindent
In this section we show that the four estimates of the observed photometric
luminosity
function listed in Table~1 are consistent with each other and we combine these
to our best estimate of the Malmquist uncorrected photometric luminosity
function, ${\overline\Psi}_{\rm phot}^*$. We do not make
use of the photometric luminosity function presented by Tinney (1993) in this
section. We need to obtain an independent estimate of the photometric
luminosity function
because in Section~5 it is our aim to test the
accuracy of the photometric luminosity function used by Tinney (1993).

\nobreak\vskip 10pt\nobreak
\noindent{\bf 3.1 Tests}
\nobreak\vskip 10pt\nobreak
\noindent
Consider the hypothesis $H_1$ that the four photometric luminosity
functions $\Psi_{\rm RG}^*, \Psi_{\rm GRH}^*, \Psi_{\rm SIP}^*, \Psi_{\rm
Kirk}^*$
(Table~1) estimate the same underlying parent distribution $\Psi_{\rm phot}^*$.
We test this hypothesis with the chi-squared test
and compute the reduced chi-squared

$$\chi_\nu^2 =
{1\over\nu}\sum_{i=1}^{I}{\left(\Psi_{1,i}^*-\Psi_{2,i}^*\right)^2
\over\sigma_{1,i}^{*2}+\sigma_{2,i}^{*2}}, \eqno (1)$$

\noindent where $\Psi_{1,i}^*$ and $\Psi_{2,i}^*$ are the luminosity function
estimates under consideration in magnitude bin $i$
and $\sigma^*_{1,i}$ and $\sigma^*_{2,i}$ are their respective
uncertainties; $I$ is the number of magnitude bins, and in our present test the
number of degrees of freedom is $\nu=I$ (no parameters are estimated from the
samples).

The significance probability $p_{\chi^2}=P(\chi^2\ge \chi_\nu^2)$ of obtaining
a chi-squared value as large as or larger than $\chi_\nu^2$
is the measure of the level of significance of the result.
We take as the significance probability at which we reject the
hypothes is $\alpha_{\rm r}=0.01$,
which corresponds to obtaining $\chi^2\ge\chi_\nu^2$ once in a hundred samples.
If $p_{\chi^2}<\alpha_{\rm r}$ then we are safe to discard $H_1$.

We test the hypothesis on both sides of the maximum in the stellar luminosity
function, as well as over the whole range, and tabulate $\chi_\nu^2$ and
$p_{\chi^2}$ in Table~3.
Since $p_{\chi^2}>\alpha_{\rm r}$ for all tests in Table~3 we do not reject
$H_1$.

The chi-squared test is valid if the underlying population
distribution is approximately normal. In order to
verify our conclusion we apply the non-parametric
Wilcoxon signed-rank test, which we discuss in greater detail in Section~4.2.
Applying this test on all pairs $(\Psi_k^*,\Psi_l^*)$, where $k,l=$RG,GRH,SIP,
Kirk, we obtain significance probabilities
$p_{T^+}=P(T^+\ge x)>\alpha_{\rm r}$ in all cases, where $T^+$ is the sum of
the ranks.
Details are tabulated in Table~4. The number of nonzero differences is $n'$.
On the basis of the Wilcoxon signed-rank test we
cannot reject $H_1$.

However, the inability to tell the luminosity functions
apart is not equivalent to showing they are alike. The average velocity
dispersion
of the stellar population in the Galactic disk is about
50~km~sec$^{-1}$. Over a length scale of 400~pc the stellar population will
thus mix within about 8~Myrs. Since the Sun is located within a few tens pc of
the Galactic midplane we thus have no astrophysical reason to {\it expect}
significantly different low-mass stellar populations along different lines of
sight.

We conclude that it is
safe to assume that $\Psi_{\rm RG}^*,\Psi_{\rm GRH}^*,\Psi_{\rm SIP}^*,
\Psi_{\rm Kirk}^*$ can
be used to compute an improved estimate of the Malmquist uncorrected
photometric luminosity function.

\nobreak\vskip 10pt\nobreak
\noindent{\bf 3.2 The best estimate of the parent raw photometric luminosity
function}
\nobreak\vskip 10pt\nobreak
\noindent
We compute the weighted average observed photometric luminosity
function

$$\overline{\Psi}_{\rm phot}^* = {\sum_i w_i \Psi_i^* \over \sum_i w_i}, \eqno
(2a)$$

$$ \overline{\sigma}^* = \left(\sum_i w_i \right)^{-{1\over2}},   \eqno (2b) $$

\noindent where $i=$RG,GRH,SIP,Kirk and $w_i=1/\sigma_i^{*2}$.  It is tabulated
in Columns~10 and~11 (where we tabulate $n_{\rm b}$, which is the sum of all
stars in each magnitude bin, rather than
$\overline{\sigma}^*$) in Table~1 and plotted in Fig.~1.

\nobreak\vskip 10pt\nobreak
\noindent{\bf 3.3 Malmquist corrections}
\nobreak\vskip 10pt\nobreak
\noindent
In Section~2 we have outlined the necessity of correcting the observed
photometric luminosity function for Malmquist bias. The Malmquist correction is
a function of $M_{\rm V}$ and can be written as

$$\Delta\Psi = \Psi_{\rm phot}^* - \Psi_{\rm phot}
= \sigma_{\rm tot}^2\, {\cal F}, \eqno (3)$$

\noindent where
${\cal F}={1\over2}\left[(0.6\,{\rm ln}10)^2\Psi_{\rm phot}
                 -(1.2\,{\rm ln}10) {d\Psi_{\rm phot}\over dM_{\rm V}}
                   + {d^2\Psi_{\rm phot}\over dM_{\rm V}^2}\right] $
(equation~17 in Stobie et al. 1989, cf also with equation~11 in Kirkpatrick et
al. 1994). The four estimates of $\Psi_{\rm phot}^*$ we use were obtained
using the same photometric bands, and can be assumed to stem from the same
parent distribution which we estimate to be ${\overline \Psi}_{\rm phot}^*$ in
Section~3.2. We assume that the true parent distribution, $\Psi_{\rm phot}$,
can be approximated by the same Gaussian model as used by Stobie et al. (1989)
to evaluate ${\cal F}$. We will verify this below by (i) showing in Fig.~3 that
our best estimate
Malmquist corrected photometric luminosity function, ${\overline\Psi}_{\rm
phot}$, is indistinguishable from the Malmquist corrected photometric
luminosity function estimated by Stobie et al. (1989), and (ii) by deriving the
(incorrect) Malmquist corrections applied by Kirkpatrick et al. (1994),
$\Delta\Psi_{\rm Kirk}(M_{\rm V})$, from our Malmquist corrections,
$\Delta\Psi(M_{\rm V})$.

Our assumption implies $\Delta\Psi=\Delta\Psi_{\rm SIP}$, which are the
Malmquist corrections computed by Stobie et al. (1989). These we plot in the
upper panel of Fig.~2, where we demonstrate that (i) the uncertainties,
${\overline\sigma}^*$, in our best estimate observed photometric luminosity
function
are significantly smaller than the Malmquist corrections, $\Delta\Psi$, for
$11.5<M_{\rm V}<14$, (ii) $\sigma_{\rm
Kirk}^*\approx\sigma_{\rm SIP}^*$ because both samples have approximately the
same number of stars per magnitude bin, (iii) the Malmquist corrections
applied by Krikpatrick et al. (1994) significantly underestimate the Malmquist
bias, and (iv)
$\Delta\Psi_{\rm Kirk}^{\rm est}\approx
({{\overline\sigma}_{\rm M}\over0.51})^2\,\Delta\Psi_{\rm SIP}$
is a good estimate of
$\Delta\Psi_{\rm Kirk}$, where ${\overline\sigma}_{\rm M}$ is the uncertainty
in $M_{\rm V}$ used by Kirkpatrick et al. (1994) to compute $\Delta\Psi_{\rm
Kirk}$ and is tabulated in their table~7. The correct cosmic scatter is
$\sigma_{\rm tot}\approx0.51$~mag (Stobie et al. 1989, Kroupa et al. 1993).

Point~(iv) verifies that our assumption of a universal ${\cal F}$ for
photometric luminosity functions obtained in the V- and I-bands
is reasonable because Kirkpatrick et al. (1994) estimate ${\cal F}$
using $\Psi_{\rm Kirk}^*$ instead of an assumed model $\Psi_{\rm phot}$ for the
parent Malmquist corrected photometric luminosity function.

For comparison with $\Delta\Psi(M_{\rm V})$ we plot in the lower panel of
Fig.~2 $\Delta\Psi_{\rm TRM}(M_{\rm bol})$, which are the first order Malmquist
corrections estimated for the photometric luminosity function in bolometric
magnitudes by Tinney et al. (1993). Clearly $\Delta\Psi_{\rm
TRM}>>\Delta\Psi$. We also observe that
the uncertainties, $\sigma_{\rm TRM}^*$, in the Tinney et al. (1993)
survey are significantly better than $\sigma_{\rm SIP}^*$ and $\sigma_{\rm
Kirk}^*$. That is, the {\it precision} of the estimate of the photometric
luminosity function is improved significantly by the survey of Tinney et al.
(1993).

We obtain our best estimate, ${\overline\Psi}_{\rm phot}$, of the Malmquist
corrected photometric luminosity function in the V-band by evaluating
${\overline\Psi}_{\rm phot}={\overline\Psi}_{\rm phot}^*-\Delta\Psi_{\rm
phot}$, with ${\overline\sigma}={\overline\Psi}_{\rm phot}/n_{\rm
b}^{1\over2}$, where $n_{\rm
b}$ are the number of stars in each magnitude bin obtained by adding the number
of stars in each survey used to estimate the photometric luminosity function
(Table~1). Our best estimate for the true distribution of stars with $M_{\rm
V}$, as obtained from deep surveys, is tabulated in Table~2.

\vskip 24pt
\noindent{\bf 4 ARE THEY DIFFERENT?}
\vskip 12pt
\noindent
In this section we compare the nearby luminosity function, $\Psi_{\rm near}$,
with the best estimate Malmquist corrected photometric luminosity function,
${\overline\Psi}_{\rm phot}$.

We note from Table~2 and Fig.~1 that the data in the nearby sample are sparse
at $M_{\rm V}>13$. In
this situation it is common practice to group the data into larger intervals to
reduce loss of information (see e.g. Bhattacharyya \& Johnson 1977, p.18).
We combine the $M_{\rm V}=14,15$ and
$M_{\rm V}=16,17$ bins to obtain $\Psi_{\rm
near,sm}=(13.9\pm5.9)\times10^{-3}$pc$^{-3}$mag$^{-1}$ in the first two bins,
and $\Psi_{\rm near,sm}=(11.4\pm5.4)\times10^{-3}$pc$^{-3}$mag$^{-1}$ in the
last two bins and plot these in Fig.~3.

{}From the figure it is apparent that the nearby luminosity
function differs at faint magnitudes from $\overline{\Psi}_{\rm phot}(M_{\rm
V})$, but that both agree approximately at bright magnitudes. In this section
we assess the {\it significance} of the difference between the nearby and
photometric luminosity functions evident to the eye-ball in Fig.3.

Consider the hypothesis $H_2$ that $\Psi_{\rm near}(M_{\rm V})$ and
$\overline{\Psi}_{\rm phot}(M_{\rm V})$ estimate the same underlying stellar
luminosity function. If the difference suggested in Fig.~3 between the two
luminosity functions is a chance fluctuation (Reid 1987, 1991)
then $H_2$ cannot be rejected at the level of significance $\alpha_{\rm r}$
(Section~3.1).

To ensure statistical rigour
three tests are applied: the chi-squared test, the Wilcoxon signed-rank
test and estimation of the probability of observing the number of stars as are
in the different samples.

\nobreak\vskip 10pt\nobreak
\noindent{\bf 4.1 The chi-squared test}
\nobreak\vskip 10pt\nobreak
\noindent
The large Poisson uncertainties in each magnitude bin in the nearby luminosity
function
preclude comparing the shapes of the two luminosity functions in detail.
We therefore combine the $M_{\rm V}=9-12$ ($i=1-4$) and the $M_{\rm V}=13-16$
($i=1-4$)
bins and compute for $\Psi_{\rm near}$ ($j=1$) and $\overline{\Psi}_{\rm phot}$
($j=2$)

$$\Psi_{{\rm tot},j} = \sum_{i=1}^4\Psi_{i,j},   \eqno (4a)$$

$$\sigma_{{\rm tot},j} = \left(\sum_{i=1}^4\sigma_{i,j}^2\right)^{1\over2},
\eqno (4b)$$

\noindent
and

$$\chi_{\rm near}^2 = {\left(\Psi_{{\rm tot},1} - \Psi_{{\rm tot},2}\right)^2
            \over
            \sigma_{{\rm tot},1}^2 + \sigma_{{\rm tot},2}^2 }. \eqno (5)$$

\noindent In Table~5 we tabulate $\chi_{\rm near}^2$ for the bright and faint
ends, and the significance probability $p_{\chi^2}=P(\chi^2\ge \chi_{\rm
near}^2)$.

As evident from the entries in Table~5 the chi-squared test implies that the
hypothesis, H$_2$, that both the nearby and the photometric
luminosity functions estimate the same distribution of stars with $9\le M_{\rm
V}\le12$ cannot be rejected. However, the hypothesis can be rejected with
high confidence ($p_{\chi^2}<\alpha_{\rm r}$) for stars with $13\le M_{\rm
V}\le 16$.

\nobreak\vskip 10pt\nobreak
\noindent{\bf 4.2 The Wilcoxon signed-rank test}
\nobreak\vskip 10pt\nobreak
\noindent
As mentioned in Section~3.1 the chi-squared test can be applied safely if the
underlying population distribution is approximately normal. Because we seek to
make a rigorous comparison between the photometric and nearby luminosity
functions we also apply a distribution-free, or nonparametric test. The
Wilcoxon signed-rank test (see e.g. Bhattacharyya \& Johnson 1977, p.519) is
designed
to allow rigorous comparison between two data sets which can consist of a
small ($n'\le15$) number of data points.
This test requires no
assumption about the shape of the population distribution, and allows for the
magnitude of the difference between the data in two samples.
In our case the data points in one sample
are ($\Psi,M_{\rm V}$).

We compute the difference $\Psi_{\rm near}-\overline{\Psi}_{\rm phot}$
in each magnitude bin, $M_{\rm V}=9,10,...16$,
and assign a rank to each of the differences which are ordered according to
increasing absolute value. The sum of the ranks of the
positive differences is $x$, and the sample size is $n'=8$. The significance
probability of obtaining a signed-rank statistic $T^+$ as large as or larger
than $x$ is
$p_{T^+}=P(T^+\ge x)$ which is evaluated from tables. The result is listed
in Table~6.

We conclude that hypothesis $H_2$ can be rejected safely because
$p_{T^+}<\alpha_{\rm r}$.

\nobreak\vskip 10pt\nobreak
\noindent{\bf 4.3 The Gau{\ss} test}
\nobreak\vskip 10pt\nobreak
\noindent
In Sections~4.1 and~4.2 we have seen that we have strong confidence that
$\Psi_{\rm near}$ does not stem from the same parent distribution as
$\overline{\Psi}_{\rm phot}$. Nevertheless, we apply another test here which
is
based on estimating the probability of observing $n$ stars in the photographic
(i.e. low spatial resolution) survey if $N$ stars have been counted in the
nearby sample.

In Table~7 we list the details. Column~1 contains the $M_{\rm V}$ bins on which
the number of stars is based. We have combined the magnitude bins in which the
stellar number density is derived from the same sampling volume $V$ in the
nearby luminosity function. Column~2 lists the volume
$V$ of the nearby sample,
and the number of stars $N$ observed in this volume is listed in Column~3.
The number of stars expected to be counted in $V$ in the photographic survey is
$n=V\,\sum_{M_{\rm V}}\overline{\Psi}_{\rm phot}$ and is listed in Column~4.
Column~5 contains $z=(N-n)/N^{1\over2}$, and Column~6 lists the probability
$p_{\rm G}$ that
$n$ or fewer stars are observed given that the population consists of $N$
stars, and is evaluated from the standard normal distribution.

{}From Table~7 we observe that the number of stars in the photographic
survey expected to be seen in the same sampling volume as is available for the
nearby sample is consistent with the number of stars seen in the nearby sample
for $M_{\rm V}=10-11$. However, the photographic survey yields far too few
stars with $M_{\rm V}=13-16$. The probability of observing $n=5$ stars given
that we expect $N=20$ stars is $p_{\rm G}<<\alpha_{\rm r}$.

We can reject hypothesis $H_2$ with high confidence.

\bigskip
\bigbreak
\noindent{\bf 5 THE BOLOMETRIC LUMINOSITY FUNCTIONS}
\nobreak\vskip 10pt\nobreak
\noindent
The recent {\it precise} luminosity function, $\Psi_{\rm TRM}^*$, constructed
from a large-scale R- and I-band
photographic survey by Tinney et al. (1993) is tabulated in table~4 of Tinney
(1993) in bolometric magnitudes.  Comparing this bolometric
luminosity function with the nearby luminosity function he concludes that there
is no difference between the two distributions on the bolometric magnitude
scale.

In view of our discordant finding in Section~4 from Tinney's, we now
transform
our best estimate of the photometric luminosity function to the bolometric
magnitude scale. We use equation~2 of Reid (1991) to transform our $M_{\rm V}$
magnitudes to $M_{\rm K}$, and the bolometric correction given by Tinney (1993)
in his fig.~5 and obtain

$$M_{\rm V} = -5.07 + 1.73\,M_{\rm Bol}, \eqno (6a)$$

\noindent and

$$M_{\rm K}=-1.87 + 0.91\,M_{\rm Bol}. \eqno (6b)$$

\noindent This is for the present purpose of camparing our
photometric luminosity function with Tinney's an adequate
approximation in the range $5<M_{\rm K}<11$, i.e. $8.1<M_{\rm V}<17.7$, as
can be verified by consulting Reid (1991) and Tinney (1993).

The luminosity function in bolometric magnitudes is given by $\Psi(M_{\rm
Bol})=\Psi(M_{\rm V})\,dM_{\rm V}/dM_{\rm bol}$.
Our best estimate of the Malmquist corrected photometric luminosity function is
tabulated in bolometric magnitudes in Table~8, which also includes the nearby
luminosity function. Both are obtained from the data in Table~2. Column~1
contains the bolometric magnitude, and Columns~2
and~3 list the best estimate photometric luminosity function in bolometric
magnitudes. Columns~4 and~5 contain the bolometric nearby luminosity function.
The last four bins have been smoothed as described in Section~4.

In Fig.~4 we compare our best estimate Malmquist corrected photometric
luminosity
function in bolometric magnitudes with that used by Tinney (1993). It is
immediately apparent that Tinney's luminosity function has
approximately a factor of
two more stars in it than our best-estimate Malmquist corrected photometric
luminosity function. This demonstrates that Malmquist bias significantly
affects the bolometric luminosity function used by Tinney (1993).

It is
because of this that Tinney interprets his fig.~4a to mean that there is no
difference between the nearby and his luminosity function. In his fig.~4a
Tinney (1993) compares $\Psi_{\rm TRM}^*$ with the nearby luminosity function,
$\Psi_{\rm WJK}$, from Wielen et al. (1983) (transformed to $M_{\rm bol}$ by
Reid 1987) and finds they are indistinguishable at $M_{\rm bol}>9.50$. However,
in his fig.~4a there is a
significant excess of bright stars in his luminosity function w.r.t. the nearby
sample.
The number density of stars he derives at $M_{\rm
bol}=9.0$ is $\Psi_{\rm Tin}=(27.0\pm1.4)\times10^{-3}$pc$^{-3}$mag$^{-1}$.
The nearby luminosity function has at $M_{\rm
bol}=8.96$, $\Psi_{\rm WJK}=(16.4\pm2.9)\times10^{-3}$pc$^{-3}$mag$^{-1}$.
The difference corresponds to $\chi^2=10.8$ which, if both samples are assumed
to estimate the same stellar population, has a probability of occurrence of
0.001. Incompleteness of the nearby sample because of ``serious completeness
problems'' for distances equal to or larger than 10~pc (Tinney 1993) cannot be
the
reason for the apparent deficit of the nearby stellar number density at $M_{\rm
bol}\approx9, M_{\rm V}\approx11$, as fig.~1 of Jahreiss (1994) demonstrates.
We note here that the luminosity function constructed by
Henry \& McCarthy (1990) from the star-count data within 5.2~pc
appears in fig.~4a of Tinney (1993) with a
significantly smaller number density at
$M_{\rm Bol}>9$, $\Psi_{\rm HM}\approx0.007$~pc$^{-3}$~mag$^{-1}$,
than the nearby luminosity function, which is based on the same data and has
$\Psi_{\rm WJK}\approx0.022$~pc$^{-3}$~mag$^{-1}$. This is due to an incorrect
scaling of $\Psi_{\rm HM}$ in his fig.~4a (Tinney, private communication).

In Fig.~5 we plot
the first order Malmquist corrected Tinney luminosity function (Tinney et al.
1993), and find good agreement with our best estimate photometric
luminosity function at $M_{\rm Bol}>9.5$. However, at brighter magnitudes
significant disagreement persists. Because the Malmquist corrections have been
applied to first order only we can at present only take the corrected Tinney
luminosity function shown in Fig.~5 to be suggestive rather than conclusive.

In Fig.~5 we also plot the
nearby luminosity function, $\Psi_{\rm near}$, in bolometric magnitudes
(Table~8) which we compare
to the nearby luminosity function, $\Psi_{\rm WJK}$, tabulated by Reid (1987).
Our transformation
of $\Psi_{\rm near}$ to $M_{\rm bol}$ agrees with that of Reid
although he used different bolometric corrections. We note here that in Fig.~5
we have corrected the erroneous entry in table~1 of Reid (1987) at $M_{\rm
V}=13, M_{\rm Bol}=10.17$. Reid lists $\Psi_{\rm
near}=22.1\times10^{-3}$pc$^{-3}$mag$^{-1}$, which has been adopted by Tinney
(1993) in his fig.~4a. The correct value is $\Psi_{\rm
near}=30.4\times10^{-3}$pc$^{-3}$mag$^{-1}$ which follows from $\Psi_{\rm
near}(M_{\rm Bol})=\Psi_{\rm near}(M_{\rm V})\,dM_{\rm V}/dM_{\rm Bol}$ and
Reid's equation~5a.

We observe from Fig.~5 that (i) our best-estimate photometric luminosity
function and the nearby luminosity function agree at $M_{\rm Bol}<9.5$, (ii)
the best-estimate photometric luminosity function and the nearby luminosity
function are significantly different at $M_{\rm Bol}>10$,
and (iii) the first order Malmquist corrected photometric luminosity
function estimated by Tinney et al. (1993) agrees with our best estimate,
$\overline{\Psi}_{\rm phot}(M_{\rm Bol})$ for $M_{\rm bol}>9.8$. Point (ii) is,
naturally, the same difference as already apparent in the photometric V-band
(Section~4).

\bigskip
\bigbreak
\noindent{\bf 6 CONCLUSIONS}
\nobreak\vskip 10pt\nobreak
\noindent
The photographic surveys using V- and I-band photometry by
Reid \& Gilmore (1982), Gilmore et al. (1985), Stobie et al.
(1989) and the CCD survey by Kirkpatrick et al. (1994) yield photometric
luminosity functions
which can be assumed to estimate the same underlying apparent distribution of
stars with luminosity. We estimate the true parent distribution by evaluating a
weighted average  Malmquist corrected
photometric luminosity function, $\overline{\Psi}_{\rm phot}(M_{\rm V})$,
which
we tabulate in Table~2 in the V-band and in Table~8 in bolometric magnitudes.

We compare our best estimate of the true Malmquist corrected photometric
luminosity function, $\overline{\Psi}_{\rm phot}(M_{\rm V})$, with the
nearby luminosity function using three different tests and conclude with a
confidence probability of at least 0.99
that the two observational approaches estimate different distributions.
We find substantial evidence that the photometric luminosity function contains
significantly fewer stars at faint magnitudes ($M_{\rm V}>13$) than the nearby
luminosity function.
In other words, the hypothesis that the best estimate Malmquist
corrected photometric luminosity
function is the correct single star luminosity function and that the nearby
luminosity function is a random fluctuation of this luminosity function (Reid
1987, 1991) can be rejected safely. Rather, the nature of the stellar
population must be different at $M_{\rm V}>13$.  Varying the Galactic disk
scale height does not affect these conclusions
provided the vertical Galactic disk scale height $h>250$~pc approximately.

The difference between
the nearby and the photometric luminosity functions
persists in bolometric magnitudes. Tinney's (1993) conclusion that
they do not differ is based on a comparison of the nearby luminosity
function with his
photometric luminosity function which is not corrected for Malmquist bias and
which therefore does not describe the distribution of stars with luminosity
{\it accurately}.

\bigskip
\bigbreak
\par\noindent{\bf ACKNOWLEDGMENTS}
\nobreak
I thank T. Henry and T. Simon for carefully reading the manuscript and many
suggestions which improved the presentation.
I also thank H. Bernstein, G. Gilmore, N. Reid, C. Tinney and C. A. Tout
for useful and much appreciated comments.

\vfill\eject

\bigbreak
\vskip 3mm
\bigbreak

\noindent
{\bf Table 1.} Estimates of the stellar luminosity function
($\times 10^{-3}$~pc$^{-3}$ mag$^{-1}$)

\nobreak
\vskip 1mm
\nobreak
{\hsize 13 cm \settabs 26 \columns

\+$M_{\rm V}$
&&&$\Psi_{\rm RG}^*$   &&$n_{\rm b}$
&&&$\Psi_{\rm GRH}^*$  &&$n_{\rm b}$
&&&$\Psi_{\rm SIP}^*$  &&$n_{\rm b}$
&&&$\Psi_{\rm Kirk}^*$ &&$n_{\rm b}$
&&&$\overline{\Psi}_{\rm phot}^*$ &&$n_{\rm b}$
\cr

\+8    &&&...  &&...&&&...  &&...  &&&4.571&&6  &&&...  &&...
&&&4.571&&6\cr

\+9    &&&2.754&&3  &&&2.239&&2  &&&4.074&&15 &&&...  &&...
&&&3.336&&20\cr

\+10   &&&6.918&&10 &&&4.677&&6  &&&6.026&&16 &&&...  &&...
&&&5.827&&32\cr

\+11   &&&9.772&&14 &&&5.754&&8  &&&9.333&&33 &&&9.669&&27
&&&8.663&&82\cr

\+12   &&&19.95&&29 &&&11.48&&16 &&&14.13&&51 &&&16.42&&46
&&&14.91&&142\cr

\+13   &&&9.120&&13 &&&15.49&&22 &&&10.00&&37 &&&12.64&&35
&&&11.13&&107\cr

\+14   &&&6.918&&10 &&&4.467&&6  &&&4.266&&16 &&&2.512&&7
&&&3.732&&39\cr

\+15   &&&3.467&&5  &&&2.344&&3  &&&2.570&&3  &&&2.188&&6
&&&2.484&&17\cr

\+16   &&&1.622&&1  &&&1.585&&1  &&&2.692&&1  &&&...  &&...
&&&1.768&&3\cr

\+17   &&&...  &&...&&&...  &&...&&&...  &&...&&&...  &&...
&&&...&&...    \cr

}
\bigbreak\vskip 3mm

\bigbreak
\vskip 3mm
\bigbreak

\noindent
{\bf Table 2.} The best estimate Malmquist corrected photometric
luminosity function and the nearby luminosity function
($\times 10^{-3}$~pc$^{-3}$ mag$^{-1}$)

\nobreak
\vskip 1mm
\nobreak
{\hsize 13 cm \settabs 7 \columns

\+$M_{\rm V}$
&&$\overline{\Psi}_{\rm phot}$ &$n_{\rm b}$
&&$\Psi_{\rm near}$ &$n_{\rm b}$ \cr

\+5  &&...   &...    &&3.223&108\cr

\+6  &&...   &...    &&3.611&121\cr

\+7  &&...   &...    &&3.044&102\cr

\+8  &&3.802 &~~6    &&3.939&~99\cr

\+9  &&2.573 &~20    &&4.745&119\cr

\+10 &&5.296 &~32    &&7.311&~23\cr

\+11 &&9.102 &~82    &&10.18&~32\cr

\+12 &&12.01 &142    &&17.71&~~7\cr

\+13 &&6.750 &107    &&12.65&~~5\cr

\+14 &&2.417 &~39    &&5.061&~~2\cr

\+15 &&1.956 &~17    &&22.77&~~9\cr

\+16 &&1.214 &~~3    &&10.12&~~4\cr

\+17 &&...   &...    &&12.65&~~5\cr

}
\bigbreak\vskip 3mm

\bigbreak
\vskip 3mm
\bigbreak

\noindent
{\bf Table 3.} Chi-squared and significance probability.

\nobreak
\vskip 1mm
\nobreak
{\hsize 16 cm \settabs 37 \columns

\+
&&&~$\Psi_{\rm RG}^*,\Psi_{\rm GRH}^*$
&&&&&&~$\Psi_{\rm RG}^*,\Psi_{\rm SIP}^*$
&&&&&&~$\Psi_{\rm RG}^*,\Psi_{\rm Kirk}^*$
&&&&&&~$\Psi_{\rm GRH}^*,\Psi_{\rm SIP}^*$
&&&&&&~$\Psi_{\rm GRH}^*,\Psi_{\rm Kirk}^*$
&&&&&&~$\Psi_{\rm SIP}^*,\Psi_{\rm Kirk}^*$
\cr

\+~$M_{\rm V}$
&&&$\nu$ &~$\chi_\nu^2$ &&~$p_{\chi^2}$
&&&$\nu$ &~$\chi_\nu^2$ &&~$p_{\chi^2}$
&&&$\nu$ &~$\chi_\nu^2$ &&~$p_{\chi^2}$
&&&$\nu$ &~$\chi_\nu^2$ &&~$p_{\chi^2}$
&&&$\nu$ &~$\chi_\nu^2$ &&~$p_{\chi^2}$
&&&$\nu$ &~$\chi_\nu^2$ &&~$p_{\chi^2}$
\cr

\+~9--12
&&&4 &1.35  &&0.25
&&&4 &0.63  &&0.57
&&&2 &0.32  &&0.75
&&&4 &0.93  &&0.45
&&&2 &1.87  &&0.15
&&&2 &0.28  &&0.78
\cr
\+13--16
&&&4 &0.85  &&0.48
&&&4 &0.39  &&0.72
&&&3 &1.69  &&0.22
&&&4 &0.59  &&0.68
&&&3 &0.48  &&0.70
&&&3 &0.84  &&0.48
\cr
\+~9--16
&&&8 &1.10  &&0.35
&&&8 &0.51  &&0.85
&&&5 &1.14  &&0.35
&&&8 &0.76  &&0.65
&&&5 &1.04  &&0.40
&&&5 &0.62  &&0.69
\cr
}
\bigbreak\vskip 3mm

\bigbreak
\vskip 3mm
\bigbreak

\noindent
{\bf Table 4.} Wilcoxon signed-rank test

\nobreak
\vskip 1mm
\nobreak
{\hsize 15 cm \settabs 28 \columns

\+~$\Psi_{\rm RG}^*,\Psi_{\rm GRH}^*$
&&&&&~$\Psi_{\rm RG}^*,\Psi_{\rm SIP}^*$
&&&&&~$\Psi_{\rm RG}^*,\Psi_{\rm Kirk}^*$
&&&&&~$\Psi_{\rm GRH}^*,\Psi_{\rm SIP}^*$
&&&&&~$\Psi_{\rm GRH}^*,\Psi_{\rm Kirk}^*$
&&&&&~$\Psi_{\rm SIP}^*,\Psi_{\rm Kirk}^*$
\cr

\+$n'$  &$x$ &~~$p_{T^+}$
&&&$n'$ &$x$ &~~$p_{T^+}$
&&&$n'$ &$x$ &~~$p_{T^+}$
&&&$n'$ &$x$ &~~$p_{T^+}$
&&&$n'$ &$x$ &~~$p_{T^+}$
&&&$n'$ &$x$ &~~$p_{T^+}$
\cr

\+8  &28 &0.098
&&&8 &22 &$>0.13$
&&&5 &11 &$>0.16$
&&&8 &26 &$>0.13$
&&&5 &5  &$>0.16$
&&&5 &9  &$>0.16$
\cr

}
\bigbreak\vskip 3mm

\bigbreak
\vskip 3mm
\bigbreak

\noindent
{\bf Table 5.} Chi-squared test
($\Psi_{\rm near},{\overline\Psi}_{\rm phot}$)
\nobreak
\vskip 1mm
\nobreak
{\hsize 5 cm \settabs 3 \columns

\+$M_{\rm V}$ &$\chi_{\rm near}^2$ &$p_{\chi^2}$\cr
\+~9--12       &2.23 &$0.15$\cr
\+13--16      &11.31 &$<0.001$\cr
}

\bigbreak\vskip 3mm

\bigbreak
\vskip 3mm
\bigbreak

{\hang{\bf Table 6.} Wilcoxon signed-rank test
($\Psi_{\rm near},{\overline\Psi}_{\rm phot}$)}

\nobreak
\vskip 1mm
\nobreak
{\hsize 5 cm \settabs 3 \columns

\+&$n'$ &$x$ &$p_{T^+}$ \cr
\+&8 &35 &0.008 \cr

}
\bigbreak\vskip 3mm

\bigbreak
\vskip 3mm
\bigbreak

{\hang {\bf Table 7.}  The number of stars $n$ expected in the photopraphic
surveys}

\nobreak
\vskip 1mm
\nobreak
{\hsize 13 cm \settabs 7 \columns
\+&$M_{\rm V}$ &$V [{\rm pc}^3]$  &$N$ &$n$ &$z$  &$p_{\rm G}$ \cr
\+&10--11 &3143.4 &55 &45 &1.31 &0.19\cr
\+&13--16 &395.2  &20 &5  &3.38 &0.0008 \cr

}
\bigbreak\vskip 3mm

\bigbreak
\vskip 3mm
\bigbreak

{\hang {\bf Table 8.} The photographic and nearby bolometric luminosity
functions. ($\times 10^{-3}$~pc$^{-3}$ mag$^{-1}$)}

\nobreak
\vskip 1mm
\nobreak
{\hsize 13 cm \settabs 10 \columns

\+&$M_{\rm Bol}$
&&$\overline{\Psi}_{\rm phot}$ &$\overline{\sigma}$
&&$\Psi_{\rm near}$ &$\sigma$ \cr

\+&8.12  & &4.46  &1.00  & & 8.22  & 0.75\cr
\+&8.70  & &9.17  &1.62  & & 12.66 & 2.64\cr
\+&9.27  & &15.77 &1.74  & & 17.63 & 3.12\cr
\+&9.85  & &20.80 &1.74  & & 30.68 & 11.60\cr
\+&10.43 & &11.69 &1.13  & & 21.91 & 9.80 \cr
\+&11.00 & &4.19  &0.67  & & 24.11 & 10.28\cr
\+&11.58 & &3.39  &0.82  & & 24.11 & 10.28\cr
\+&12.16 & &2.10  &1.21  & & 19.73 & 9.30\cr
\+&12.73 & &...   &...   & & 19.73 & 9.30\cr

}
\bigbreak\vskip 3mm


\vfill\eject

\bigskip
\noindent{\bf REFERENCES}
\nobreak
\bigskip
\nex Bhattacharyya, G. K., Johnson, R. A., 1977, Statistical Concepts and
     Methods, John Wiley \& Sons Press, New York
\nex Dahn, C. C., Liebert, J., Harrington, R. S., 1986, AJ 91, 621
\nex Gilmore, G., Reid, N., Hewett, P., 1985, MNRAS 213, 257
\nex Hawkins, M. R. S., Bessell, M. S., 1988, MNRAS 234, 177
\nex Haywood, M., 1994, A\&A 282, 444
\nex Henry, T. J., McCarthy, D. W., 1990, ApJ 350, 334
\nex Jahreiss, H., 1994, Ap\&SS 217, 63
\nex Kirkpatrick, J. D., McGraw, J. T., Hess, T. R., Liebert, J., McCarthy, D.
     W., 1994, ApJS 94, 749
\nex Kroupa, P., 1995a, Unification of the Nearby and Photometric Stellar
      Luminosity
      Functions, ApJ, in press
\nex Kroupa, P. Tout, C. A., Gilmore, G., 1991, MNRAS 251, 293
\nex Kroupa, P., Tout, C. A., Gilmore, G., 1993, MNRAS 262, 545
\nex Leggett, S. K., Hawkins, M. R. S., 1988, MNRAS 234, 1065
\nex Lutz, T. E., Kelker, D. H., 1973, PASP 85, 573
\nex Reid, N., 1987, MNRAS 225, 873
\nex Reid, N., 1991, AJ 102, 1428
\nex Reid, N., 1994, Ap\&SS 217, 57
\nex Reid, N., Gilmore, G., 1982, MNRAS 201, 73
\nex Stobie, R. S., Ishida, K., Peacock, J. A., 1989, MNRAS 238, 709
\nex Tinney, C. G., 1993, ApJ 414, 279
\nex Tinney, C. G., 1994, The Luminosity and Mass Functions at the Bottom of
     the Main Sequence, In: MacGillivray, H. T., Thomson, E. B., Lasker, B. M.,
     et al., (eds.), Astronomy from Wide-Field Imaging, Kluwer, Dordrecht,
     p.411
\nex Tinney, C. G., Reid, N., Mould, J. R., 1993, ApJ 414, 254
\nex Wielen, R., Jahreiss, H., Kr{\"u}ger, R., 1983. The Nearby Stars
     and Stellar Luminosity Function, In: Davis Philip, A.
     G., Upgren, A. R. (eds.), IAU Colloq. No. 76, New York, Davis Press, p.163

\vfill\eject


\vfill\eject

\centerline{\bf Figure captions}
\smallskip

\noindent {\bf Figure 1.} The observational data (Section~2). The photometric
luminosity
functions not corrected for Malmquist bias are shown as the long-dashed curve
(Reid \& Gilmore 1982, $\Psi_{\rm RG}^*$), as the dot-short-dashed curve
(Gilmore et al. 1985, $\Psi_{\rm GRH}^*$), as the short-dashed curve (Stobie et
al. 1989, $\Psi_{\rm SIP}^*$) and as the dot-long-dashed curve (Kirkpatrick et
al. 1994, $\Psi_{\rm Kirk}^*$). The open circles are the weighted average,
${\overline\Psi}_{\rm phot}^*$, of these four luminosity functions, and the
solid circles are our best estimate of the
Malmquist corrected photometric luminosity function, ${\overline\Psi}_{\rm
phot}$, derived in Section~3.3. The nearby luminosity function, $\Psi_{\rm
near}$, is shown as the open squares.

\vskip 5mm

\noindent {\bf Figure 2.} The Malmquist corrections and Poisson uncertainties
(Section~3.3).
{\bf Upper panel}: We compare the Malmquist correction computed by Stobie et
al.
(1989), $\Delta\Psi_{\rm SIP}$, and the Poison uncertainties in their raw
photometric luminosity function, $\sigma_{\rm SIP}^*$, with these same
quantities
computed by Kirkpatrick et al. (1994). The Poisson uncertainties of our
weighted average raw photometric luminosity function, ${\overline\sigma}^*$,
are improved substantially over the uncertainties in the individual estimates
of the photometric luminosity function (Fig.~1). {\bf Lower panel}: The first
order Malmquist corrections estimated by Tinney et al. (1993) and their Poisson
uncertainties.

\vskip 5mm

\noindent {\bf Figure 3.} The luminosity function of stars in the photometric
V-band (Section~4). The solid histogram is the nearby luminosity function
$\Psi_{\rm
near}(M_{\rm V})$. We have grouped the data into larger bins at $M_{\rm
V}\ge13.5$. The filled circles are our best-estimate
of the Malmquist corrected photometric luminosity function,
$\overline{\Psi}_{\rm phot}(M_{\rm
V})$. $\Psi_{\rm near}$ and
$\overline{\Psi}_{\rm phot}$ are tabulated in Table~2. The short dashed curve
is the Malmquist corrected photometric luminosity function estimated by Stobie
et al. (1989).

\vskip 5mm

\noindent {\bf Figure 4.} Our best-estimate photometric luminosity function in
bolometric magnitudes, $\overline{\Psi}_{\rm phot}(M_{\rm Bol})$ (filled
circles, Table~8), is
compared with the luminosity function used by Tinney (1993), $\Psi_{\rm
TRM}^*(M_{\rm Bol})$ (open triangles) (Section~5).

\vskip 5mm

\noindent {\bf Figure 5.} The solid histogram is the nearby bolometric
luminosity function, and the filled circles are our best-estimate
Malmquist corrected photometric luminosity function (Section~5). The
luminosity function
used by Tinney (1993) and plotted in Fig.~4 has been corrected here to first
order for Malmquist bias (Tinney et al. 1993). It is shown as open triangles.
The open stars are the bolometric luminosity function derived from the
nearby star sample by Reid (1987), after correcting his erroneous value at
$M_{\rm Bol}=10.17$.

\vfill
\bye